\documentclass[prl,twocolumn,preprintnumbers]{revtex4}

\usepackage{graphicx,amsmath,amsfonts,textcomp}

\newcommand{\fref}[1]{Fig.~\ref{#1}}

\begin{document}
\title{Nanoscale polarization switching mechanisms in multiferroic BiFeO$_3$ thin films\\
}

\author{H. B\'ea}
\email{helene.bea@unige.ch}
\affiliation{DPMC, University of Geneva, 24 Quai Ernest Ansermet,
1211 Geneva 4, Switzerland}
\author{M. Bibes}
\affiliation{Unit\'e Mixte de Physique CNRS/Thales, 1 Avenue A. Fresnel, 91767 Palaiseau, France and Universit\'e Paris-Sud, 91405 Orsay, France}
\author{A. Barth\'el\'emy}
\affiliation{Unit\'e Mixte de Physique CNRS/Thales, 1 Avenue A. Fresnel, 91767 Palaiseau, France and Universit\'e Paris-Sud, 91405 Orsay, France}
\author{P. Paruch}
\affiliation{DPMC, University of Geneva, 24 Quai Ernest Ansermet,
1211 Geneva 4, Switzerland}
\date{\today}
\begin{abstract}
Ferroelectric switching in BiFeO$_3$ multiferroic thin films with intrinsic ``stripe-like'' and ``bubble-like'' polydomain configurations was
studied by
piezoresponse force microscopy.  Using the local electric field applied by a scanning probe microscope tip, we observe reversal of both
out-of-plane
and in-plane components of the polarization, with the final domain state depending on the tip sweeping direction.  In ``bubble-like'' samples,
complete
control of the polarization is achieved, with in-plane polarization change mediated and stabilized by out-of-plane polarization reversal.
In
``stripe-like'' samples the intrinsic domain structure influences polarization switching and in-plane reversal may occur without out-of-plane
change.  The observed switching behaviour can be well correlated with the radial and vertical components of the highly inhomogeneous electric
field applied by the tip.
\end{abstract}
\maketitle

The co-existence of ferroelectric and antiferromagnetic ordering at room temperature in BiFeO$_3$ (BFO) has made this material the focus of
intense research,
both aimed at understanding its fundamental properties, and at exploiting them in multifunctional device applications (see for instance
\cite{Catalan09,Bibes08} and references therein).
Magnetoelectric coupling between the orders, allowing the rotation of the antiferromagnetic ordering via the application of electric field,
has
been demonstrated in both thin films \cite{Zhao06} and bulk single crystals \cite{Lebeugle08}, although the bulk magnetoelectric coefficient
has
been determined to be extremely small \cite{Popov93}.  The magnetoelectric effect was subsequently used to control the exchange bias \cite{BeaEB}
and to
electrically control the magnetization of metallic layers magnetically coupled to BFO  \cite{Chu08}. In spite of these application-oriented
advances, open questions remain about the mechanism behind the observed magnetoelectric coupling, and especially about its
effects at the nanoscale, the proposed regime for future device implementation.  Extensive quantitative studies have been rendered challenging
in
part by the complex ferroelectric structure of BFO, with eight equivalent variants of polarization giving rise to nanoscale domains separated
 by
three different types of domain walls (71, 109 and 180$^{\circ}$ \cite{Zhao06}), some of which themselves present additional functionalities,
such as
conduction \cite{Seidel09}, absent from the parent material.  Moreover, in epitaxial thin films, different intrinsic ferroelectric domain
configurations can be observed, in which some of the eight polarization variants appear to be suppressed depending on the
growth conditions, the bottom electrode, the substrate and substrate miscut, ranging from monodomain, to ``stripe-like'' and
``mosaic-like'' polydomain \cite{Chu07} and to ``bubble-like'' polydomain \cite{Catalan08}.
For fundamental insight into the mechanisms behind magnetoelectric coupling in BFO thin films, and for their use in technological applications,
nanoscale
control of the ferroelectric domain structure and understanding of local ferroelectric switching mechanisms is thus of key importance.
Macroscopic polarization switching has been studied using parallel electrodes to apply homogenous out-of-plane  \cite{Gruverman05} or in-plane
electric fields \cite{Shafer07}, albeit separately. To be able to both switch the polarization and to measure all its components, it is also possible
to use
liquid electrodes \cite{Rodriguez07}. Another option to study full polarization switching and to avoid leakage is to use the metallic tip of
an atomic force microscope as an electrode \cite{Zavaliche05APL}. In this configuration, a local, intense, but highly inhomogeneous field is
generated with
both a vertical component
allowing out-of-plane switching, and a horizontal component with approximately radial symmetry due to the tip shape. After switching,  piezoresponse
force microscopy (PFM) can be used to fully characterize the polarization, allowing access to nanoscale switching mechanisms.

In this letter, we present a detailed PFM study of the polarization switching mechanisms that occur in BFO thin films under a biased atomic
force
microscope tip.  In the case of ``bubble-like'' BFO films,  we observe switching of the in-plane polarization only when associated with out-of-plane
polarization reversal.  In the case of ``stripe-like'' BFO films,  the in-plane
component of the polarization may switch independently. In both cases the radial electric field of the tip, as well as its sweep direction, strongly
influence the polarization switching.
In ``bubble-like'' samples, these mechanisms may allow a complete control of both components of the polarization without in-plane electrodes.

For these studies, we used 50--70 nm thick BFO films deposited on (001)-oriented SrTiO$_3$ (STO) substrates by pulsed
laser deposition \cite{BeaAPL2} (giving ``bubble-like'' films) and rf magnetron sputtering (``stripe-like''), on top of 25-35 nm thick  SrRuO$_3$,
used as a bottom electrode.
The high crystalline quality of the samples has been confirmed by X-ray diffraction (see \cite{Beaphilmag} for the 70nm ``bubble-like'' sample,
similar results being obtained for the 50 nm ``stripe-like'' film). The rms roughnesses of the presented films are 2nm in the ``bubble-like' case and 0.4nm
in the ``stripe-like'' case. Ferroelectric domain writing and PFM
measurements were performed using a {\it Nanoscope V Dimension} (20kHz, 3-4V ac, NCS18/Cr-Au from $\mu${\it masch}). In these films, we wrote domain
structures by
applying a voltage (equivalent to a nominal field of around 170MV/m in both presented samples)
to the conductive tip while sweeping it in the desired pattern. Because the polarization in BFO lies close to the $<$111$>$ directions, and thus
presents both out-of-plane and in-plane components in these (001) oriented films, the written regions are then measured by both vertical and lateral PFM
(VPFM and
LPFM, respectively). When discussing PFM images, we mean PFM phase image, unless otherwise stated.

\begin{figure}
\includegraphics[width=0.9\columnwidth]{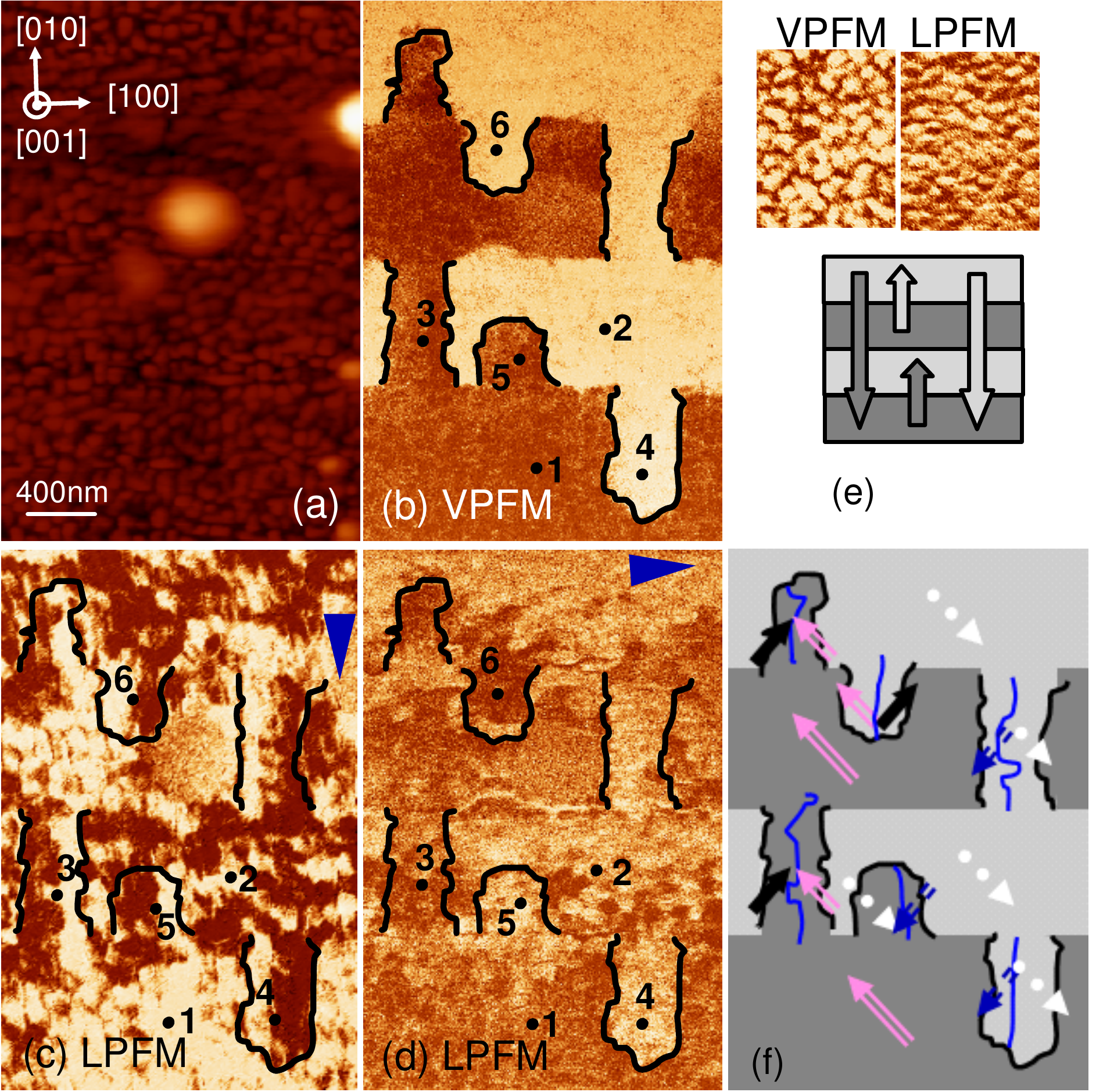}
\caption{``Bubble-like'' film: (a) Topography, (b) VPFM, (c) and (d)  LPFM after writing large horizontal rectangles with negative and positive
voltages and then sweeping the tip upwards or downwards. The blue triangles indicate the orientation of the cantilever during LPFM measurements.
As-grown VPFM and LPFM images and the written pattern are indicated in upper part of (e). Lower part of (e): dark gray corresponds to -12V, light gray
to +12V, and the arrows indicates
the sweeping direction. For the rectangles, the slow scan axis was along [0$\overline{1}$0]. The black lines correspond to domain walls written during
vertical tip sweeps (b). (f) Out-of-plane (bright and dark colors) and in-plane orientation of polarization (arrows).
The black lines follow those of (b). In the case of the non-monodomain rectangular region (see c), only the predominant in-plane component is reported.
For the numbers, see text.}
\label{BFO001}
\end{figure}

We first studied ``bubble-like'' polydomain BFO films presenting all eight variants of the polarization in the as-grown regions, in which VPFM
and LPFM show an intrinsic domain structure somewhat correlated with surface morphology (\cite{Catalan08} and \fref{BFO001}e). On such film, we
prepolarized wide horizontal lines with alternating positive and negative voltage
on a
3x3$\mu$m$^2$ area.
Subsequently, narrower vertical lines were written with positive or negative voltage while sweeping the tip either along [010] or [0$\overline{1}$0]
(\fref{BFO001}e). The domains written with negative voltage (``up'' polarization, regions 1, 3, 5) show a dark
VPFM contrast (\fref{BFO001}b), while the positive-voltage-written domains are bright (regions 2, 4, 6). As expected, when a vertical line written with
negative (resp. positive) voltage crosses a region prepolarized with positive (resp. negative) voltage, the out-of-plane polarization component is
reversed, while when it crosses a negatively (resp. positively) prepolarized region no change in the VPFM contrast is observed.

In the LPFM images (\fref{BFO001}c and d), we observe that the in-plane configuration of the polarization has also been clearly modified as compared
to the as-grown film (\fref{BFO001}e). In negatively
prepolarized rectangles (region 1), the LPFM signal of \fref{BFO001}c presents a bright contrast indicating an in-plane component along [$\overline{1}$00], while
for the positively prepolarized rectangles (region 2), a predominant dark contrast is obtained with a certain amount of bright regions, giving a
predominant component along [100]. In these rectangles, the LPFM signal along the [010] direction (\fref{BFO001}d) shows a bright
contrast after a positive applied voltage (region 2) and a dark contrast after a negative applied voltage (region 1). In the negatively polarized
rectangles, we thus obtain a monodomain configuration \footnote{The as-grown configuration of this film being preferentially ``down'' (\fref{BFO001}e),
in the positively prepolarized rectangles, the in-plane configuration was not fully controlled due to the need for an accompanying out-of-plane
switching}.

During the vertical tip sweeps, the in-plane component of the polarization was similarly modified, but
only in the specific regions in which the out-of-plane polarization component  was also reversed during writing (outlined in black in \fref{BFO001}b-d).
In these regions, for the cantilever oriented along the vertical lines (\fref{BFO001}c), a characteristic pattern is observed: a dark LPFM contrast on
the right of the cantilever and a bright one on its left for the negative-voltage-written lines (regions 3, 5) while the reverse is true
for lines written with a positive voltage (regions 4, 6). This  pattern correlates with the horizontal component of the electric field perpendicular
to the cantilever
axis. The in-plane polarization switching is thus assisted by the
out-of-plane polarization
switching  in this range of fields and reproduces the horizontal electric field distribution.

Imaging the in-plane domain structure of the vertical lines with the cantilever rotated by 90$^{\circ}$ (\fref{BFO001}d), we again observe that the only
regions where the in-plane component of the polarization has been modified are the regions where the out-of-plane component of the polarization has also
been reversed (again, outlined in black). In these measurements, however, the LPFM contrast not only depends on the polarity of the applied voltage
but also varies as a function of the tip sweeping direction during writing. For a negative voltage, where the tip moved along [0$\overline{1}$0]
(region 3), a dark contrast is observed whereas where the tip moved along [010], the contrast is bright (region 5). For a positive voltage, the opposite is
observed: where the tip moved along [0$\overline{1}$0], the contrast is bright (region 4) and when the tip moved along [010], the contrast is dark
(region 6).
In \fref{BFO001}f, we have summarized both out-of-plane (bright or light gray) and in-plane polarization orientations (arrows) in the written regions.

\begin{figure}
\includegraphics[width=0.9\columnwidth]{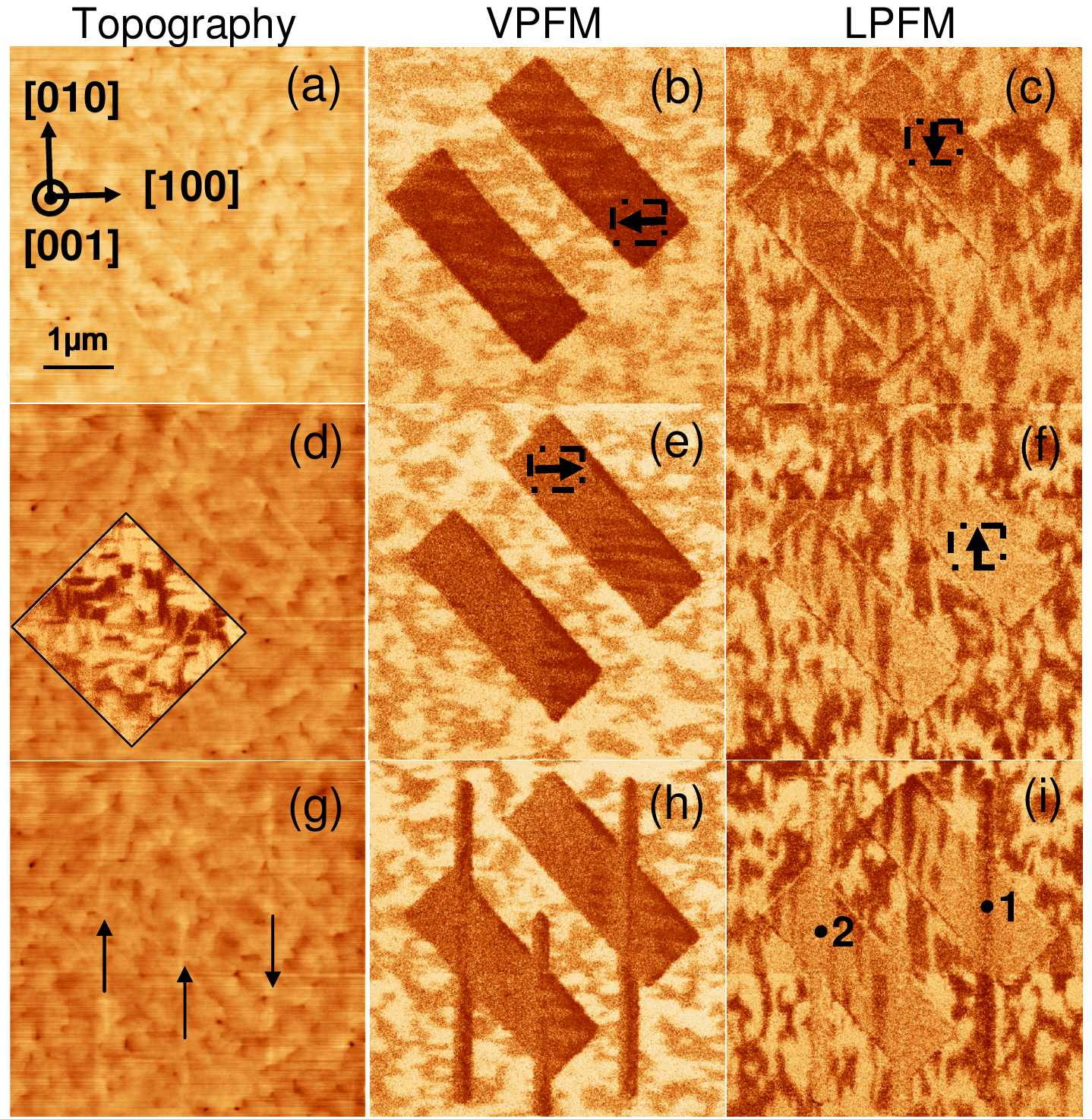}
\caption{``Stripe-like'' film: Topography (a), (d), (g), VPFM (b), (e), (h) and LPFM (c), (f), (i) images taken with the cantilever along the
[$\overline{1}$00] direction after writing different domain patterns. Two rectangles were written with negative voltage by sweeping the tip at
45$^{\circ}$ with the slow scan axis along [$\overline{1}\overline{1}$0] (a-c) or [110] (d-f). (g-i): vertical lines were written with a
negative-voltage tip sweeping in the direction given by the arrow in (g) on top of (d-f). The in-plane direction of polarization along [100]
(b and e) and [010] (c and f), corresponding to the observed color contrast is indicated by arrows for a given region. The inset of (d) is a
LPFM image of an as-grown region measured with the cantilever along [110]. For the numbers, see text.}
\label{BFOsputt}
\end{figure}

We then carried out similar measurement on ``stripe-like'' films, as shown in \fref{BFOsputt}.
In this case, the as-grown film presents only the four ``down'' polarization components (ie. bright contrast in  VPFM images of
\fref{BFOsputt}b, e, h), with stripe-like features as observed in the LPFM image in inset of \fref{BFOsputt}d. We note that the VPFM
contrast  is not homogeneously bright in the as-grown region, but presents clear structure, with faintly darker regions corresponding to a noisier
phase and to a decrease in VPFM amplitude (data not shown). This may be attributed to a buckling of the cantilever in response to an in-plane
piezoresponse along the cantilever axis \cite{Peter06,Shafer07}. Contrary to the ``bubble-like'' film \footnote{In the
``bubble-like'' film, the roughness was higher, probably making the buckling signal more difficult to access.}, in the case of a ``stripe-like''
film, only one VPFM image and one LPFM image are thus necessary to reconstruct the total polarization configuration.

On this sample, we have written two sets of two rectangles with negative voltage by sweeping the tip with its slow scan axis along either
[$\overline{1}\overline{1}$0] (\fref{BFOsputt}a-c)
or [110]  (\fref{BFOsputt}d-f). In both cases, the out-of-plane polarization has been switched in these regions, as checked by a measurement
made with the cantilever rotated by 90$^{\circ}$ to distinguish deflection from buckling contributions (not shown).

Considering the in-plane signals, we see that in the LPFM images (\fref{BFOsputt}c and f) the written regions are not fully monodomain,
but rather present a predominant contrast with a few stripes corresponding to the opposite in-plane polarization orientation.
Similarly, in these regions, the VPFM signal is not homogeneous due to the buckling contribution, and is composed in \fref{BFOsputt}b (resp.
\fref{BFOsputt}e) of a predominant dark (resp. faintly brighter) contrast with a few stripes presenting the opposite contrast, ie. faintly
brighter (resp. dark). Comparing the sets of measurements of \fref{BFOsputt}b-c and e-f, we clearly see that these predominant contrasts are
opposite, thus revealing opposite in-plane component of the polarization both along [100] (see boxes and arrows in \fref{BFOsputt}b and e)
and [010] directions (see boxes and arrows in \fref{BFOsputt}c and f). To summarize, the predominant resulting in-plane configuration is reversed
when the slow scan axis during writing was along [110] or [$\overline{1}\overline{1}$0], in agreement with the conclusions obtained for the
``bubble-like'' film.

As for the ``bubble-like'' film, we then wrote vertical lines with negative tip voltage, sweeping along the directions indicated in \fref{BFOsputt}g,
and measured the resulting domain pattern (\fref{BFOsputt}g-i). Strikingly, and unlike what we observe in the ``bubble-like'' film, the in-plane
 component of the polarization has been modified even when the out-of-plane polarization was not switched. This can be seen in particular in
 \fref{BFOsputt}i, where a dark (region 1) or bright (region 2) line appears in the previously written rectangle. Also, the in-plane contrast of the written lines along
 [100] (from buckling signal in \fref{BFOsputt}h) is similar to that obtained for the negative-voltage-written lines in the ``bubble-like''
 film, ie. directed towards the tip. Along the [010] direction (from LPFM in \fref{BFOsputt}i),
 the  contrast depends on the sweeping direction: a bright contrast is obtained for the tip sweeping along [010] (region 2) and a dark contrast for the
 [0$\overline{1}$0] direction (region 1).

To understand the observed switching of the in-plane polarization components, we need to consider the  electric field generated by the AFM tip.
\begin{figure}
\includegraphics[width=0.9\columnwidth]{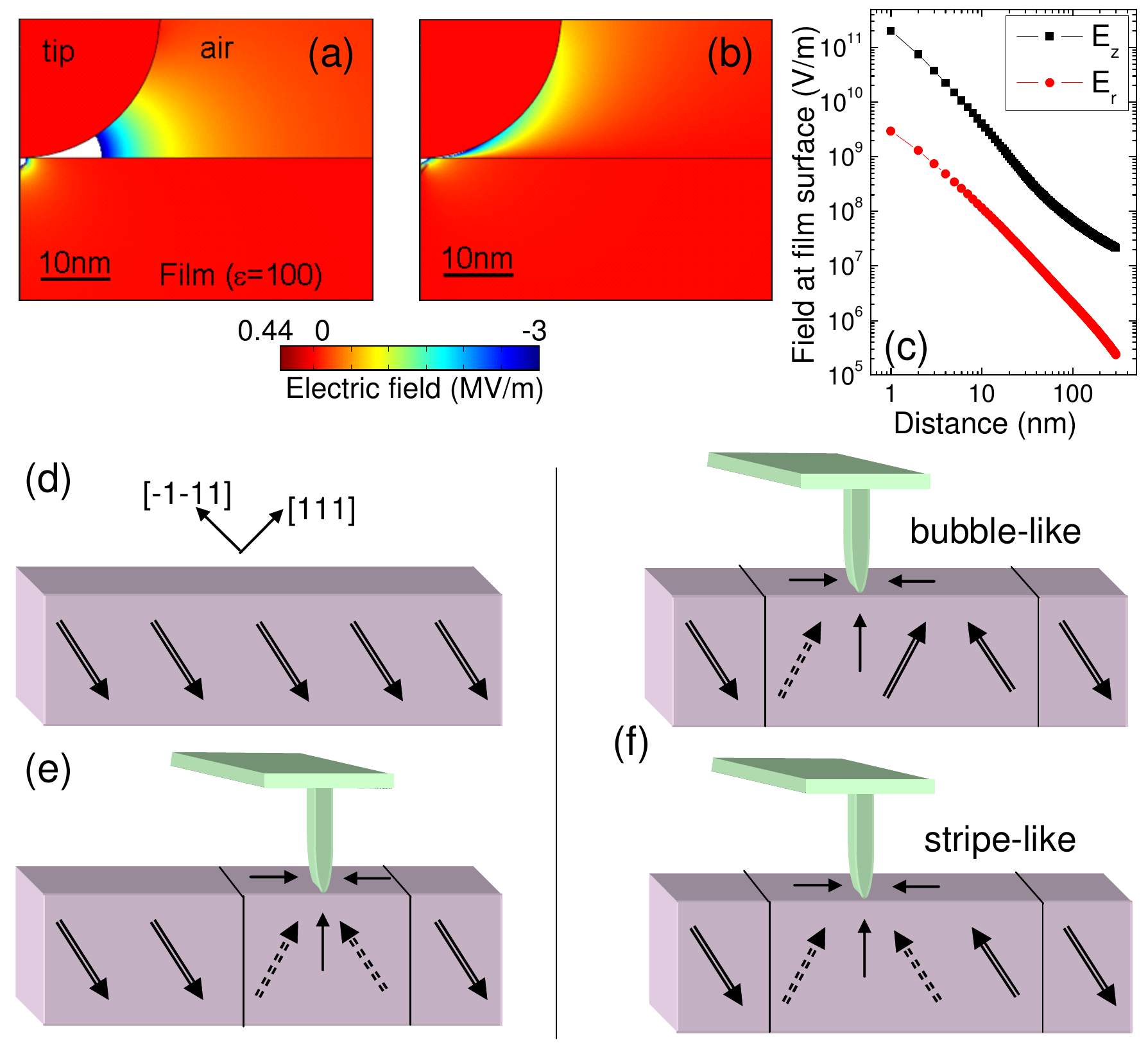}
\caption{(a) Vertical (E$_z$) and (b) horizontal (E$_r$) electrical field distribution when 12V is applied to the tip. The color scales for the field
being the same for (a) and (b), the white regions correspond
to out-of-range field values. (c) Log-Log plot of E$_z$ and E$_r$ at the film surface as a function of the
horizontal distance from tip apex.  (d)-(f) Schematic representation of the switching mechanisms: first (d), the polarization is homogeneous
(wide arrows). Under a negative tip voltage, the polarization switches (e) just below the tip (dashed wide arrows), following the direction of
both vertical and horizontal field (represented by the small narrow arrows). Then, when the tip sweeps leftwards (f) with the same negative
voltage, in the case of the ``bubble-like'' films (upper part), the in-plane polarization
will be modified only if the out-of-plane component of the polarization is switched, while in the ``stripe-like'' case (lower part) it may
switch independently.}
\label{comsolmech}
\end{figure}
Using {\it COMSOL Multiphysics}, we modeled a tip with a realistic shape in contact with a 200nm-thick insulating film  presenting a dielectric
constant of 100 on top of a metal electrode. As expected,
this electric field  is  highly inhomogeneous with both a vertical component E$_z$ (\fref{comsolmech}a) and a radially-symmetric horizontal
component E$_r$ (\fref{comsolmech}b). On the film surface, as plotted in
\fref{comsolmech}c, both vertical and horizontal components of the electric
field decrease as the inverse of the distance from the tip. However their magnitudes are very different, the horizontal field being two orders
of magnitude smaller than the vertical field.  In fact, simply considering the electric field profile, switching the in-plane polarization using
the AFM tip would appear to be significantly more difficult than switching the out-of-plane polarization.

Coherently with this picture, we previously observed that in (111)-oriented films presenting two purely out-of-plane polarization variants, the
application of an electric field through the tip results only  in the purely out-of-plane variant \cite{BeaAPL111}, in agreement with other
experiments \cite{Seidel09}. In that case where a purely out-of-plane polarization is possible, the in-plane field is thus too small to rotate
the polarization in a strongly in-plane direction, and a purely out-of-plane polarization is stabilized during switching. On the other hand,
in the case of (001)-oriented films, all variants of the polarization have equivalent in-plane components. The small in-plane field of the AFM
tip thus renders these variants inequivalent, and could lead to stabilization of a modified in-plane polarization component.

To summarize these observations, we propose a switching mechanism for the polarization in a monodomain BFO thin film under a biased tip, as
presented in \fref{comsolmech}d-f. These figures give a schematic representation of a negatively polarized tip sweeping towards the left. First,
for both types of samples, the polarization initially oriented along [11$\overline{1}$] (\fref{comsolmech}d) is switched underneath the tip: on
the left of the tip, the polarization is now oriented along [111] while on its right, it is oriented along [$\overline{1}$$\overline{1}$1]
(\fref{comsolmech}e),
following the vertical  and horizontal local electric field. Then, in a ``bubble-like'' film,
as the tip sweeps leftwards (\fref{comsolmech}f, upper part), switching will continue only on the left of the cantilever: the region on its
right being already ``up'', neither the out-of-plane nor the in-plane component of the polarization will change, in agreement with experimental
observations. In the switching region, the horizontal field of the tip is directed right, and the polarization will now point in the [111]
direction. In the end, the in-plane component of the polarization will thus present a thin stripe with polarization oriented along
[$\overline{1}$$\overline{1}$1] in the initial switching region and then along [111] for the rest of the area swept by the tip. For a
``stripe-like'' film, when the tip sweeps leftwards (\fref{comsolmech}f, lower part), both in-plane and out-of-plane components of the polarization
may be switched: the region on the right of the tip in the [$\overline{1}\overline{1}$1] direction, while on its left in the [111] direction.
Obviously, in both types of samples, if the tip sweep direction or voltage are  changed, the final in-plane component of the polarization will
be affected. In this representation of switching, we therefore expect the final in-plane component of the polarization to be oppositely directed
for the same writing pattern for the two different types of film. This expectation is borne out experimentally: for negative voltage, in the
``bubble-like'' films, the predominant in-plane component of the polarization is opposite to the sweeping direction (see regions 1, 3, 5 of \fref{BFO001}) while in the ``stripe-like''
films, it is along the sweeping direction (see boxes and arrows in \fref{BFOsputt}).

To conclude, we have shown the possibility of a full control over the $<$111$>$-oriented polarization direction of ``bubble-like'' (001)-BFO
thin films by using a biased sweeping AFM tip. In ``stripe-like'' films, an incomplete in-plane switching and non-monodomain regions are obtained,
depending on pre-existing domain structure.  In these latter films, the final stable domain structure imaged by PFM could in fact be the result of
 a more complex switching path exploring the eight variant configurations  \cite{BalkePFM09}.
The switching mechanism in the two types of films appears to be different, with in-plane switching occurring more easily and independently of
out-of-plane switching in the ``stripe-like'' films. This suggests that the coercive fields for in-plane and out-of-plane components of the
polarization could be different in the two cases, which could be linked to surface morphology, type of domain walls or pinning defects.
Different behavior depending on the as-grown domain and domain wall configuration of the BFO thin films was already observed
for instance in exchange biased
ferromagnetic/BFO  bilayers \cite{MartinEB08}.
Finally, the polarization control in ``bubble-like'' films demonstrated in this work, allowing desired domains and domain
walls in BFO thin films to be defined using small in-plane fields, can be very useful in future devices and for fundamental studies of
magnetoelectric coupling mechanisms at the nanoscale.

\begin{acknowledgments}
The authors thank J.-M. Triscone, K. Bouzehouane and S. Fusil for helpful discussions. Technical support by M. Lopes and E. Jacquet is acknowledged. This work was
supported by the Swiss National Science Foundation through the NCCR MaNEP and Division
II, by the European Commission STREP project MaCoMuFi. H.B. is supported by a Bourse d'Excellence from the University of Geneva.
\end{acknowledgments}

\end{document}